\documentstyle[11pt,a4]{article}

\newcommand{\rf}[1]{(\ref{#1})}
\newcommand{\beq}{\begin{equation}}
\newcommand{\eeq}{\end{equation}}
\renewcommand{\c}{\gamma}
\renewcommand{\b}{\beta}
\renewcommand{\a}{\alpha}

\renewcommand{\d}{\delta}

\renewcommand{\l}{\lambda}
\newcommand{\bea}{\begin{eqnarray}}
\newcommand{\eea}{\end{eqnarray}}
\newcommand{\cD}{{\cal D}}
\newcommand{\cQ}{{\cal Q}}

\newcommand{\ex}{\uparrow}
\newcommand{\re}{\downarrow}
\newcommand{\fr}{\frac}

\begin{document}

\title{Ultrametric matrices and representation theory}
\author{
P. B\'antay and G. Zala \\
{\em \scriptsize Institute for Theoretical Physics, 
E\"otv\"os University, H-1088 Puskin u. 5-7, Budapest, Hungary}
}
\date{\today}
\maketitle

\begin{center}
{\sc Abstract} 
\end{center}
{\em \small The consequences of replica-symmetry breaking on the structure of 
ultrametric 
matrices appearing in the theory of disordered systems is investigated with 
the help of representation theory, and the results
are compared with those obtained by Temesv\'ari, De Dominicis and Kondor. 
}

\begin{center}
\end{center}

\section{Introduction}

The technique of replica-symmetry breaking provides a general framework to
describe the microscopic properties of low-temperature disordered systems. 
Originally developed in the theory of spin glasses \cite{parisi}, this method 
had found applications in a wide variety of problems, 
such as the theory of random manifolds \cite{random1,random2,random3},
vortex pinning \cite{vortex}, random-field problems \cite{rfield1,rfield2}, etc.
In these theories randomness is handled via the replica trick, and the multitude
of equilibrium phases is captured by breaking the permutation symmetry 
between replicas. 

In the replica method the free-energy $F=F(q_{\a \b})$ depends on a set of 
order parameters $q_{\a \b}$, where the replica indices $\a,\b$ 
take integer values in the set $\{1, 2,\ldots ,n\} $ and the order parameter 
matrix is symmetric with  zero diagonal entries. The free-energy is 
independent of the labeling of the replicas, i.e. $F$ is invariant under the 
transformations $q_{\a \b} \to q_{\pi(\a) \pi(\b)}$ for $\pi \in S_n $, where
$S_n$ denotes the symmetric group of rank $n$, that is the group of all 
permutations of the integers $\{1,2, \ldots, n \}$. 

Depending on the value of the parameters in $F$,  
the stationary points of the free-energy are either symmetrical, meaning 
that all of their off-diagonal components are equal,
or replica-symmetry breaking. 
As usual, symmetry breaking means that the ground state is invariant only
under a proper subgroup of the underlying $S_n$ symmetry group of the theory.
Many important features
of the theory follow from the residual symmetry of the ground state by
standard arguments based on the Wigner-Eckart theorem.

The successful Ansatz for the 
symmetry breaking pattern, first proposed by Parisi \cite{parisi}, 
looks as follows. Let $R$ be a 
positive integer and let the positive integers $p_0=n,p_1, p_2, \ldots ,p_R$ be 
such that $p_{i+1}$ divides $p_i$.
The $n \times n$ 
matrix $q_{\a \b}$ is divided into blocks of size $p_1 \times p_1$, and a 
common value $q_0$ is assigned to all matrix elements outside the diagonal 
blocks. 
Next, the diagonal blocks are further divided into blocks of size 
$p_2 \times p_2$ and the value $q_1\ne q_0$ is assigned to all elements 
inside the 
diagonal blocks of size $p_1 \times p_1$ but outside the diagonal blocks of 
size $p_2 \times p_2$, and so on down to the innermost blocks of size 
$p_R \times p_R$, where the matrix elements are $q_R$, except the diagonal of 
the whole matrix where they are all zero. 

The residual symmetry group is by definition that subgroup of $S_n$ which leaves
the saddle-point invariant, i.e.
\begin{equation}
H_{(p_0,p_1,\ldots,p_R)}= 
   \{ \pi \in S_n | \ q_{\pi(\a) \pi(\b)} = q_{\a \b} \ \mbox{for all} \ 
   \a, \b = 1,2,\ldots,n \}. 
\end{equation}
The structure of this group for a Parisi-type saddle-point is captured by 
the notion of a wreath product of symmetric groups \cite{robinson}. 
Let $k$ be a divisor of 
$n$ and divide the natural numbers $\{ 1,2, \ldots ,n\}$ 
into $l$ $(=n/k)$ blocks of length $k$. 
The wreath product $S_k \wr S_l$ is the group of permutations 
which move blocks as a whole with permutations from $S_l$, and also the 
elements inside the blocks with permutations from $S_k$. 
Generalization to multiple wreath products is obvious and it can 
be shown that the wreath product is associative, i.e. brackets can be omitted.  
The residual symmetry group at a Parisi-type saddle-point, which we shall 
denote by $H$ in the following, is isomorphic to the multiple wreath product
\begin{equation}
S_{p_R} \wr S_{p_{R-1}/p_{R}} \wr \ldots \wr S_{p_0 / p_1}.
\end{equation}

The second derivative $ M_{\a \b, \c \d}:={\partial^2 F}/{\partial q_{\a \b} 
\partial q_{\c \d}} $ of the free energy evaluated at a Parisi-type saddle-point
is a four-index quantity  whose special properties following from the 
symmetry breaking pattern are usually referred to as ultrametricity. 
The characterization of a generic
ultrametric matrix - block-diagonalization, spectral decomposition - was 
given in~\cite{kondor}, where it was shown that there exists a basis such 
that the operator $M$ is block-diagonal, containing only blocks of sizes 
$R+1$ and $1$. It was the desire to understand the group theoretic origin
of this result that led us to the present representation theoretic study of
ultrametricity. Clearly, the advantage of the group theoretic analysis is
that it may be readily generalized to more complex situations, e.g. the
study of higher rank ultrametric operators (i.e. higher derivatives of
the free-energy), whose properties are of prime interest for a better
understanding of the underlying physical theories.

\section{Ultrametric matrices}

Let us denote by $\cal Q$ the space of order parameters, i.e. the linear space
of symmetric $n \times n$ matrices with zeros on the main diagonal. The 
free-energy is a real-valued function on $\cQ$, invariant under the action 
$q_{\a \b} \to q_{\pi(\a) \pi(\b)}$ of the symmetric group $S_n$. The 
above action realizes a linear representation $D$ of $S_n$ on the space $\cQ$, 
whose decomposition into irreducibles is given by ($n \geq 5$) \cite{kerber} 
\begin{equation}
D=[n-2,2] \oplus [n-1,1] \oplus [n],
\label{D}
\end{equation}
where we have used the usual labeling of the irreps of $S_n$ via
partitions of $n$. We see that only three irreducible 
components appear, which - by analogy with the representations of the general 
linear group -  may be termed as the "tensor", "vector" and "scalar", 
respectively. 

The Hessian $M=\partial^2 F / \partial q^2$ may be viewed as a 
linear operator $M : \cQ\to\cQ$. When evaluated at a Parisi-type 
saddle-point with residual symmetry
group $H$, the invariance of $F$ with respect to the action of $S_n$ implies
$D(\pi) M D^{-1}(\pi) = M$, in other words 
\begin{equation}
[D(\pi),M]=0
\label{eq:comm}
\end{equation}
for all $\pi \in H$. 
This commutation rule is the abstract algebraic expression of the ultrametricity
of $M$, and the problem is to find out the implications of this property on
the structure of the operator, e.g. the number of different eigenvalues together
with their multiplicities.

Such conclusions may be drawn by a clever application of the Wigner-Eckart
theorem. For suppose we know the decomposition into irreducibles of
the restriction $D\re H$ of the representation $D$ to the residual subgroup 
$H$ :
\begin{equation}
D\re H=\bigoplus_i m_i C^{(i)},
\end{equation} 
where the $C^{(i)}$ denote the irreps of $H$, and $m_i$ is the multiplicity
of the corresponding irrep. Then the Wigner-Eckart theorem tells us that 
in a suitable basis the ultrametric matrix $M$ is block-diagonal, having
blocks of size $m_i$ appearing with multiplicity $d_i$, equal to the dimension
of the irrep $C^{(i)}$. Moreover, the diagonal blocks may be written down
explicitly by applying suitable projection operators completely determined
by the irreps $C^{(i)}$.

\section{The decomposition of $D \re H$}

The structure of the residual subgroup $H$ and of its irreducible
representations change markedly as we increase the number $R$ of
symmetry breaking steps. It is therefore natural to try to describe
this process inductively, starting from the symmetric case where
$R=0$, and going on to the more complicated cases step-by-step.

\subsection{The $R=0$ case ($H=S_n$)}

In this case there is no symmetry breaking. 
As we have seen previously, $D$ can be decomposed into three irreps:
\( [n-2,2] \oplus [n-1,1] \oplus [n] \), which we'll denote in the sequel 
by $t_0$, $v_0$ and $s_0$ respectively, the subscript referring to the
$R=0$ case. An ultrametric  operator $M$
satisfying \rf{eq:comm} has accordingly three different eigenvalues corresponding 
to the above irreps, with respective multiplicities $\fr{1}{2} n(n-3)$,
$n-1$ and $1$.
  
\subsection{The $R=1$ case ($H=S_k \wr S_l$, $kl=n$)}
\label{r1}

We need to find the irreducible constituents of $D\re H$.
The restriction of the identity representation is trivial, but that of 
the vector and tensor requires a more 
sophisticated analysis. For details of the representation theory of
wreath products we refer to Appendix \ref{app1}.
While the proof works only for $k,l \geq 5$,
the result turns out to be valid for $k,l \geq 4$ as well.

To decompose into irreducibles the restriction to $S_k\wr S_l$ of an irrep
of $S_{kl}$, one can apply the following simple procedure :
\begin{itemize}
\item One computes the restriction of the irrep to $S_k$ by repeated 
application of the so-called branching law, which describes the
decomposition of the restriction of any irrep of $S_n$ to $S_{n-1}$.
\item From Eqs. (\ref{restr}) and (\ref{rmult}) of the Appendix 
one can compute the decomposition of any irrep of $S_k\wr S_l$ 
into irreps of $S_k$.
\item By the transitivity of restriction, the above decompositions
should agree, which constrains strongly the allowed irreducible 
constituents of the restriction to $S_k\wr S_l$.
\item If there is still some ambiguity left in the decomposition,
comparison of the character values at some specific elements will
fix the result completely.
\end{itemize}

Applying the above procedure to the restriction $v_0 \re H$ results in 
the decomposition
\beq
v_0 \re \ = v_1 \oplus v_1',\label{v0}
\eeq 
where $v_1$ and $v_1'$ are certain irreducible representations of $S_k \wr S_l$, 
to be defined in the Appendix. 
Here and from now on $\re$ denotes the restriction to the next level, i.e. 
from $H_{(p_0, p_1, \ldots, p_i)}$ to the subgroup 
$H_{(p_0, p_1, \ldots, p_i, p_{i+1})}$.
The analogous result for the tensor representation $t_0$ reads
\beq
t_0 \re \ = t_1 \oplus v_1 \oplus v_1^2 \oplus v_1 v_1' \oplus t_1' \oplus v_1' 
  \oplus s,
\label{eq:tede}
\eeq  
where $t_1$, $v_1^2$, $v_1 v_1'$ and $t_1'$ denote again irreducible
representations of $S_k \wr S_l$ to be defined in the Appendix.
Putting all together, we get in this case the result
\beq 
D \re \ = t_1 \oplus 2v_1 \oplus v_1^2 \oplus v_1 v_1' \oplus t_1' 
\oplus 2v_1'   \oplus 2s,
\eeq
i.e. a total of seven different irreducible constituents, three of them with
multiplicity 2. According to this, an ultrametric matrix has ten different
eigenvalues, whose multiplicities are determined by the dimensions of the 
above irreps (cf. Table 1).

\subsection{Generalization to $R>1$}

For a start, let's restrict the above mentioned $S_k \wr S_l$ irreps to
$(S_p \wr S_q) \wr S_l$ ($pq=k$) and decompose them. The method together
with some illustrating examples is described in Appendix \ref{app2}. 
The definitions of the irreps to appear in this subsection are
also to be found there.
For the 
decomposition of $v_1 \re$ we obtain
\beq
v_1 \re \ = v_2 \oplus v_2',
\eeq
while for $t_1 \re$ we have
\begin{equation}
t_1 \re \ =t_2 \oplus v_2 \oplus v_2^2 \oplus v_2 v_2' \oplus t_2' \oplus v_2' 
  \oplus v_1' \oplus s
\label{eq:tdec2}
\end{equation}
where the subscript $2$ refers again the $R=2$. The $R=1$ level 
representations which 
contain trivial base representations - i.e. 
$t_1'$, $v_1'$ and s - remain irreducible under restriction 
(so we denote the restricted irreps with the same symbols), while
$v_1 v_1'$ restricts simply as
\beq
v_1 v_1' \re \ = v_2 v_1' \oplus v_2' v_1'. 
\eeq
The most tricky case is the decomposition of $v_1^2\re$. The result reads
\begin{equation}
v_1^2 \re \ = v_2 \stackrel{1}{\bowtie} v_2 \oplus v_2 \stackrel{1}{\bowtie} 
v_2' \oplus v_2' \stackrel{1}{\bowtie} v_2'.
\end{equation}
To make the $R=2$ step clear, Figure 1. shows the 
decomposition tree of $t_0 \re H$ up to this level.  

%
%

\vspace{0.5cm}

\begin{figure}[h]

\setlength{\unitlength}{0.012500in}%
\begingroup\makeatletter
\def\x#1#2#3#4#5#6#7\relax{\def\x{#1#2#3#4#5#6}}%
\expandafter\x\fmtname xxxxxx\relax \def\y{splain}%
\ifx\x\y   
\gdef\SetFigFont#1#2#3{%
  \ifnum #1<17\tiny\else \ifnum #1<20\small\else
  \ifnum #1<24\normalsize\else \ifnum #1<29\large\else
  \ifnum #1<34\Large\else \ifnum #1<41\LARGE\else
     \huge\fi\fi\fi\fi\fi\fi
  \csname #3\endcsname}%
\else
\gdef\SetFigFont#1#2#3{\begingroup
  \count@#1\relax \ifnum 25<\count@\count@25\fi
  \def\x{\endgroup\@setsize\SetFigFont{#2pt}}%
  \expandafter\x
    \csname \romannumeral\the\count@ pt\expandafter\endcsname
    \csname @\romannumeral\the\count@ pt\endcsname
  \csname #3\endcsname}%
\fi
\endgroup
\begin{picture}(305,127)(95,685)
\thicklines
\put(200,740){\line(-1,-1){ 40}}
\put(200,740){\line( 0,-1){ 40}}
\put(200,740){\line( 1,-2){ 20}}
\put(200,740){\line( 1,-1){ 40}}
\put(200,740){\line(-1,-2){ 20}}
\put(200,740){\line(-3,-2){ 60}}
\put(200,740){\line( 3,-2){ 60}}
\put(200,740){\line(-2,-1){ 80}}
\put(260,800){\line( 3,-4){ 30}}
\put(290,740){\line(-1,-4){ 10}}
\put(290,740){\line( 1,-4){ 10}}
\put(260,800){\line( 3,-2){ 60}}
\put(260,800){\line( 3,-1){120}}
\put(380,740){\line(-1,-2){ 20}}
\put(380,740){\line( 3,-2){ 60}}
\put(380,740){\line( 1,-2){ 20}}
\put(330,740){\line(-1,-4){ 10}}
\put(330,740){\line( 1,-4){ 10}}
\put(260,800){\line(-5,-2){100}}
\put(260,800){\line(-3,-1){120}}
\put(260,800){\line(-4,-1){160}}
\put(260,800){\line(-3,-2){ 60}}
\put(260,805){\makebox(0,0)[lb]{\smash{\SetFigFont{12}{14.4}{rm}{\scriptsize $t_0$}}}}
\put(430,685){\makebox(0,0)[lb]{\smash{\SetFigFont{12}{14.4}{rm}{\scriptsize $v_2' \stackrel{1}{\bowtie} v_2'$}}}}
\put(135,685){\makebox(0,0)[lb]{\smash{\SetFigFont{12}{14.4}{rm}{\scriptsize $v_1'$}}}}
\put(155,685){\makebox(0,0)[lb]{\smash{\SetFigFont{12}{14.4}{rm}{\scriptsize $v_2'$}}}}
\put(170,685){\makebox(0,0)[lb]{\smash{\SetFigFont{12}{14.4}{rm}{\scriptsize $t_2'$}}}}
\put(200,685){\makebox(0,0)[lb]{\smash{\SetFigFont{12}{14.4}{rm}{\scriptsize $t_2$}}}}
\put(215,685){\makebox(0,0)[lb]{\smash{\SetFigFont{12}{14.4}{rm}{\scriptsize $v_2$}}}}
\put(230,685){\makebox(0,0)[lb]{\smash{\SetFigFont{12}{14.4}{rm}{\scriptsize $v_2 v_2'$}}}}
\put(255,685){\makebox(0,0)[lb]{\smash{\SetFigFont{12}{14.4}{rm}{\scriptsize $v_2^2$}}}}
\put(115,685){\makebox(0,0)[lb]{\smash{\SetFigFont{12}{14.4}{rm}{\scriptsize $s$}}}}
\put( 95,745){\makebox(0,0)[lb]{\smash{\SetFigFont{12}{14.4}{rm}{\scriptsize $s$}}}}
\put(135,745){\makebox(0,0)[lb]{\smash{\SetFigFont{12}{14.4}{rm}{\scriptsize $v_1'$}}}}
\put(155,745){\makebox(0,0)[lb]{\smash{\SetFigFont{12}{14.4}{rm}{\scriptsize $t_1'$}}}}
\put(195,745){\makebox(0,0)[lb]{\smash{\SetFigFont{12}{14.4}{rm}{\scriptsize $t_1$}}}}
\put(285,745){\makebox(0,0)[lb]{\smash{\SetFigFont{12}{14.4}{rm}{\scriptsize $v_1$}}}}
\put(315,745){\makebox(0,0)[lb]{\smash{\SetFigFont{12}{14.4}{rm}{\scriptsize $v_1 v_1'$}}}}
\put(375,745){\makebox(0,0)[lb]{\smash{\SetFigFont{12}{14.4}{rm}{\scriptsize $v_1^2$}}}}
\put(275,685){\makebox(0,0)[lb]{\smash{\SetFigFont{12}{14.4}{rm}{\scriptsize $v_2'$}}}}
\put(295,685){\makebox(0,0)[lb]{\smash{\SetFigFont{12}{14.4}{rm}{\scriptsize $v_2$}}}}
\put(310,685){\makebox(0,0)[lb]{\smash{\SetFigFont{12}{14.4}{rm}{\scriptsize $v_2 v_1'$}}}}
\put(332,685){\makebox(0,0)[lb]{\smash{\SetFigFont{12}{14.4}{rm}{\scriptsize $v_2' v_1'$}}}}
\put(354,685){\makebox(0,0)[lb]{\smash{\SetFigFont{12}{14.4}{rm}{\scriptsize $v_2 \stackrel{1}{\bowtie} v_2$}}}}
\put(390,685){\makebox(0,0)[lb]{\smash{\SetFigFont{12}{14.4}{rm}{\scriptsize $v_2' \stackrel{1}{\bowtie} v_2$}}}}
\end{picture}

\vspace{0.5cm}
\caption{Decomposition of $t_0$}
\end{figure}
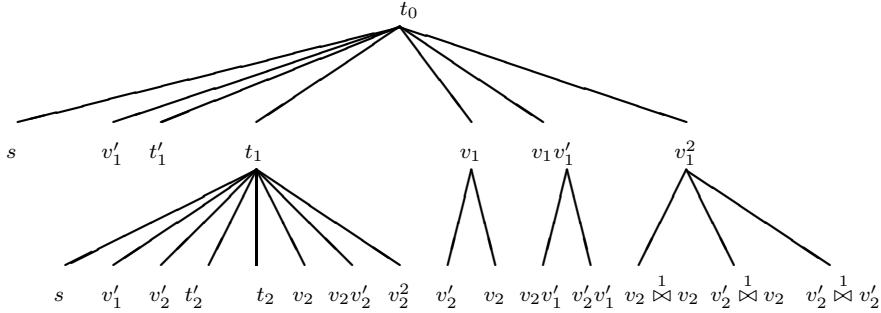
%
%
\vspace{0.5cm}

The generalization to higher $R$'s can be described inductively.
Until now to make the notation easier to understand we used $p,
q$ and $l$ instead of $p_0=n=pql$, $p_1=pq$ and $p_2=p$, but
from now on we shall proceed with the $p_i$'s ($i=1, 2, \ldots,
R$). The inductive definition of the representations to appear in
what follows is to be found in the Appendix.

$v_i\re$ splits into two representations, similarly to (\ref{v0})
\begin{equation}
v_i \re \ = v_{i+1} \oplus v_{i+1}',
\end{equation}
while $t_i\re$ splits into $i+7$ different representations:
\begin{equation}
t_i \re \ = t_{i+1} \oplus v_{i+1} \oplus v_{i+1}^2 \oplus v_{i+1} v_{i+1}' \oplus
t_{i+1}' \oplus v_{i+1}' \oplus v_{i}' \oplus \ldots \oplus v_1' \oplus s.
\end{equation}
As a general rule we can state that the inertia factor representations,
 denoted by a prime,
never decompose further (so we can denote them with the same symbol), e.g. $t_i'$ and $v_i'$ remain irreducible under
further restriction. 
$v_i v_i'$ restricts according to 
\beq
v_i v_i' \re \ = v_{i+1} v_i' \oplus v_{i+1}' v_i'.
\eeq
The above decomposition generates representations of the form $v_i v_j'$ and 
$v_i' v_j'$ with $(i \geq j)$. The restriction rule for the first type
reads
\begin{equation}
v_i v_j' \re \ = v_{i+1} v_j' \oplus v_{i+1}' v_j',
\end{equation}
while the second type remains irreducible, containing only inertia
factor representations. 
Once again the most complicated case is the decomposition of $v_i^2 \re$ 
\begin{equation}
v_i^2 \re \ = v_{i+1} \stackrel{i}{\bowtie} v_{i+1} \oplus v_{i+1} \stackrel{i}{\bowtie} 
v_{i+1}' \oplus v_{i+1}' \stackrel{i}{\bowtie} v_{i+1}'
\end{equation}
which gives birth to three new class of representations. 
\begin{equation}
v_{i+1} \stackrel{i}{\bowtie} v_{i+1} \re \ = v_{i+2} \stackrel{i}{\bowtie} v_{i+2}
\oplus v_{i+2} \stackrel{i}{\bowtie} v_{i+2}' \oplus v_{i+2}' \stackrel{i}{\bowtie} v_{i+2}'.
\end{equation}
It is now obvious that after a certain number of steps starting with $v_i^2$ we
obtain representations of the form 
\begin{equation}
v_{i} \stackrel{j}{\bowtie} v_{i} \ \ \ \ \ v_{i} \stackrel{j}{\bowtie} v_{k}'
\ \ \ \ \ v_{i}' \stackrel{j}{\bowtie} v_{k}'. 
\end{equation}
The third ones do not decompose further, the second ones split as
\begin{equation}
v_{i} \stackrel{j}{\bowtie} v_{i}' \re \ = v_{i+1} \stackrel{j}{\bowtie} v_{i}' \oplus v_{i+1}' \stackrel{j}{\bowtie} v_{i}'
\end{equation}
and the first ones split according to  
\begin{equation}
v_{i} \stackrel{j}{\bowtie} v_{i} \re \ = v_{i+1} \stackrel{j}{\bowtie} v_{i+1}
\oplus v_{i+1} \stackrel{j}{\bowtie} v_{i+1}'
\oplus v_{i+1}' \stackrel{j}{\bowtie} v_{i+1}'.
\end{equation}

Finally, after having performed $R$ reduction steps, we obtain the result 
summarized in Table 1 for the restriction of the representation $D$. 
The domain of the variables are $1 \leq i,j \leq R$ and
$1 \leq k \leq R-1$. The classification of the irreps into families
accords with that of \cite{kondor}.

{ \footnotesize
\begin{table}
\[
\begin{array}{|l|l|l|l|} \hline
\mbox{{\em Family}} & \mbox{{\em Symbol of irrep}} & \mbox{{\em Multiplicity}} 
   & \mbox{{\em Dimension}} \\ \hline 
& & & \\
L   & s & R+1 & 1 \\ \hline 
& & & \\
A   & v_i' & R+1 & n ( 1/p_i - 1/p_{i-1} ) \\ 
    & v_R  & R+1 & n(1 - 1/p_R) \\  \hline 
& & & \\
R_1 & t_i' & 1 & \fr{n}{2} (p_{i-1} - 3p_i)/p_i^2 \\ 
    & t_R & 1 & \fr{n}{2} (p_R - 3) \\  \hline 
& & & \\
R_2 & v_i' v_j'  \ ,(i>j) & 1 & n (p_{i-1}-2p_i)/p_i (1/p_j - 1/p_{j-1}) \\ 
    & v_R v_i' & 1 & n (p_{i-1}-2p_i)/p_i  (1 - 1/p_R) \\ \hline
& & & \\ 
R_3 & v_R^2 & 1 & \fr{n}{2} (p_{R-1} - p_R) (1 - 1/p_R)^2 \\ 
    & v_R \stackrel{k}{\bowtie} v_R \ & 1 
      & \fr{n}{2} (p_{j-1}-p_j) (1 - 1/p_R)^2 \\ 
    & v_i' \stackrel{k}{\bowtie} v_i' \ ,(i>k) & 1 & 
    \fr{n}{2} (p_{j-1}-p_j) (1/p_i - 1/p_{i-1})^2 \\ 
    & v_R \stackrel{k}{\bowtie} v_i' \ ,(i>k) & 1 & 
    n (p_{j-1}-p_j) (1 - 1/p_R) (1/p_i - 1/p_{i-1})  \\ 
    & v_i' \stackrel{k}{\bowtie} v_j' \ ,(i>j>k) & 1 &
    n (p_{j-1}-p_j) (1/p_k - 1/p_{k-1}) (1/p_i - 1/p_{i-1}) \\ \hline
\end{array}
\]
\caption{Irreducible constituents of $D\re H$}
\end{table}
}

\section{Discussion}

The irreps are divided into three families: $L$, $A$ and $R$ and the latter 
is subdivided 
into three subfamilies $R_1$, $R_2$ and $R_3$. The family $L$ consists of the 
trivial irrep $s$, $A$
consists of the "vector-like" irreps, and $R$ includes 
the other irreps, which are characterized by the fact that they all originate
from the tensor representation $t_0$. 
This classification accords that of \cite{kondor}. 

What kind of conclusions may be drawn from the above decomposition about
the structure of an arbitrary ultrametric matrix? As explained in section 2, 
the Wigner-Eckart theorem tells us that
the matrix may be block-diagonalized in a suitable basis. To the irreps
in the $L$ and $A$ families will correspond blocks of size $R+1$, with
multiplicities equal to the dimension of the corresponding irreps,
while the representations from the family $R$ appear only once, 
i.e. to each of them is associated a single eigenvalue of the ultrametric 
matrix, whose multiplicity is again the dimension of the corresponding irrep.
This is exactly the pattern found in \cite{kondor} - without the use of group 
theory - for the spectral decomposition of an arbitrary ultrametric matrix.

In summary, we have seen that the structure of ultrametric
matrices is to a large extent determined by the residual symmetry
group, in complete accord with the results of \cite{kondor}.
While the primary goal of the present work was to elucidate the
group theoretic background of that paper, it should be
stressed that the results may be applied in further
investigations of replica-symmetry breaking, e.g. in the
analysis of the symmetry properties of the correlation
functions. Besides this, they may lead to a better understanding
of the symmetry structures present in the physically interesting
limit $R\to\infty$, which is probably one of the most
interesting features of the theory.

\section*{Acknowledgments}
The application of representation theory techniques to the study of 
replica-symmetry breaking was pioneered by the late Claude Itzykson. 
We are grateful to I. Kondor and T. Temesv\'ari for directing our attention
to this field and for the many interesting discussions.

\appendix

\section{Representations of wreath products}
\label{app1}

First of all, we sketch briefly the representation theory 
of wreath products $G \wr S_l$ 
for a finite permutation group $G$ of degree $k$
\cite{kerber}, which is a classical application of Clifford's theorem 
\cite{isaacs}. Let's divide the natural numbers $\{1,2, \ldots, n\}$ 
into blocks of length $k$ and let $G^{(i)}$ denote the subgroup of 
$S_n$ which permutes the numbers inside the 
$i$-th block ($i=1,2, \ldots,l$). Clearly $G^{(i)} \cong G$. 
To obtain the irreps of the wreath product $G \wr S_l$ we follow the 
procedure outlined
here: 
\begin{itemize}
\item Let's first 
construct the so-called base group (containing no permutations moving whole 
blocks) 
\begin{equation}
G^{*}=G^{(1)} \times G^{(2)} \times \ldots \times G^{(l)}.
\end{equation}
The irreps of this group are of the form 
\begin{equation}
F^{(1)} \sharp F^{(2)} \sharp \ldots \sharp F^{(l)},
\label{eq:baserep}
\end{equation}
where $F^{(i)}$ is an irrep of $G^{(i)}$ and the symbol $\sharp$ denotes the
outer tensor product.
Let $D_1, D_2, \ldots, D_r$ be all the irreps of $G$ and   
define an $l$-partition $\Lambda=\langle \l_1, \l_2, \ldots, \l_r \rangle$ which 
denote the situation where $\l_1$ of the $F^{(i)}$'s are equal to $D_1$ and 
$\l_2$ of them are equal to $D_2$, etc. 
$\Lambda$ is
called the type of the base group representation and it has the property
$\sum_{i=1}^{r} \l_i = l$. 
\item Let's define the inertia factor:
\begin{equation}
S_{\Lambda}=\{ \pi \in S_l \ | \ F^{(\pi(i))}=F^{(i)} \ \mbox{for all} \ i=1,2, \ldots ,l \  \},
\end{equation}
which is isomorphic with $\times_{i=1}^{r} S_{\l_i}$.
\item Now we extend the representation from the base group to the $G \wr S_{\Lambda}$ 
inertia group:
\begin{eqnarray}
& (\overline{F_{\a_1 \b_1}^{(1)} F_{\a_2 \b_2}^{(2)} \ldots F_{\a_l \b_l}^{(i)}})
(g_1,g_2,\ldots, g_l;\sigma)= & \cr \cr
& F_{\a_1 \b_{\sigma(1)}}^{(1)}(g_1) F_{\a_2 \b_{\sigma(2)}}^{(2)}(g_2) 
\ldots 
F_{\a_l \b_{\sigma(l)}}^{(l)}(g_l) &
\end{eqnarray}
for all $g_i \in G^{(i)}$ and $\sigma \in S_{\Lambda}$.  
\item Finally the general form of an irrep $\cD$ of $G \wr S_l$ is the 
following:
\begin{equation}
(\overline{F^{(1)} \sharp F^{(2)} \sharp \ldots \sharp F^{(l)}} \otimes K) 
   \ex G \wr S_l.
\label{eq:wreathrep}
\end{equation}  
Here $K$ is an irrep of the inertia factor $S_{\Lambda}$ and hence a tensor product of
irreps of $S_{\l_i}$, i.e. $K=K_1 \sharp K_2 \sharp \ldots \sharp K_r$. 
An alternative -- shorter -- notation of (\ref{eq:wreathrep}) is
\beq
\langle D_1, K_1 \rangle \sharp \langle D_2, K_2 \rangle \sharp \ldots \sharp 
  \langle D_r, K_r \rangle .
\eeq

\end{itemize}

Let's consider the behaviour of the irrep $\cD$ under restriction. Let's single out
one of the factors of $G^*$, e.g. the first one, and consider the restriction $\cD \re
G$.
It follows from the construction that its decomposition into irreps of $G$ is given
by
\beq
\cD \re G = \bigoplus_{j=1}^r m_j D_j,
\label{restr}
\eeq
where the multiplicities $m_j$ are given by
\beq
m_j=\frac{\l_j}{l \dim(D_j)} \dim(\cD).
\label{rmult}
\eeq 
For the sake of definiteness we give the dimension of the whole
wreath product irrep $\cD$:
\beq
\dim(\cD)=l! \ \prod_{i=1}^{r} \frac{\dim(K_i) \dim(D_i)^{\l_i}}{\l_i!}
\eeq

\section{Decomposition in the $R=1$ case}
\label{app2}

We shall illustrate the procedure outlined in \ref{r1} on the decomposition of the 
vector representation $[n-1,1] \re (S_k \wr S_l)$.
The branching law tells us that
\beq [n-1,1] \re S_k = [k-1,1] \oplus k(l-1) \ [k].
\label{eq:vectbran} 
\eeq        
Now taking into account the restriction rule (\ref{restr}) and (\ref{rmult})
we conclude that the involved wreath product irreps may contain only
vector and scalar irreps in the base representation, i.e. we may deal with
a representation of the form $\langle v_0, K_1 \rangle \sharp \langle s, K_2 \rangle$. 
Since we have exactly one vector irrep 
in (\ref{eq:vectbran}), hence there must be a wreath product irrep constituent of the
decomposition with 
$\l_1 = 1$ and $\dim(K_2)=1$ ($\dim(K_2)$ denotes the dimension of the irrep
$K_2$). This implies $K_1=[1]$ and since the 
only one-dimensional irreps are the trivial and the alternating: $K_2=[l-1]$ or
$K_2=[1^l]$. 
Furthermore there must be another constituent (or other constituents) which contain no
vector irrep factor in the base representation, i.e. with  
$\l_1=0$ and $\l_2 = l$. 
The dimension of the original $v_0$ is $(kl-1)$ so the remaining dimension  
is $(l-1)$ which can be filled in several ways: we can choose 
$K_2=[l-1,1]$ or we can choose $l-1$ one-dimensional irreps with $K_2=[l]$ or $K_2=[1^l]$.

It is possible to further reduce the number of the possibilities using the
characters of the representations. Evaluating the characters of both the original and 
the candidate representation on the elements $(1 \ 2)$ and $(1 \ 2 \ 3)$ 
(permutating the
blocks) we end up with only one remaining version:
\begin{equation}
v_0 \re \ = \langle s, [l-1,1] \rangle 
  \oplus \langle v_0, [1] \rangle \sharp \langle s, [l-1] \rangle.
\label{eq:vdec}
\end{equation}

We proved that the restriction of $v_0$ to $S_k \wr S_l$ can be 
decomposed according to (\ref{eq:vdec}). This 
decomposition holds for $k,l \geq 2$.  
To simplify the notation let's introduce the symbol  
$v_1'$ for the first part of the 
decomposition and $v_1$ for the second. The
subscript $1$ at both symbol denotes the $R=1$ case. 

In case of the decomposition of $t_0 \re$ we just briefly sketch the
definitions of the resulting constituents. Since here the branching law
results [k-2,2] irreps too, the base group representations will contain 
scalar, vector and tensor as well:
\begin{eqnarray}
t_1 &=& \langle t_0, [1] \rangle \sharp \langle s, [l-1] \rangle \cr \cr 
v_1 v_1' &=& \langle v_0, [1] \rangle \sharp \langle s, [l-1,1] \rangle \cr \cr
v_1^2 &=& \langle v_0, [2] \rangle \sharp \langle s, [l-2] \rangle \cr \cr
t_1' &=& \langle s, [l-2,2] \rangle 
\label{wdef1}
\end{eqnarray}

\section{Generalization to $R>1$}
\label{app3}

Let's consider $v_1$ as an example:
\begin{eqnarray}
 & ([k-1,1]_1 \sharp [k]_2 \sharp \ldots \sharp [k]_l 
   \otimes [l-1] \ex S_k \wr S_l ) \re S_p \wr S_q \wr S_l  = & \cr \cr
 & [k-1,1]_1 \re  (S_p \wr S_q) \sharp [k]_2 \sharp \ldots 
   \sharp [k]_l \otimes [l-1] \ex S_p \wr S_q \wr S_l  &. 
\end{eqnarray}
Here we could omit the overline above the base irrep since it has no effect.
The decomposition of $[k-1,1] \re S_p \wr S_q$ is already known, so making
use of the distributivity of the tensor product we obtain the following constituents:
\begin{eqnarray}
v_2 &=& \langle v_1, [1] \rangle \sharp \langle s, [l-1] \rangle, \cr \cr
v_2' &=& \langle v_1', [1] \rangle \sharp \langle s, [l-1] \rangle
\end{eqnarray}

We have simply changed the $[k-1,1]$ factor to two different $S_p \wr S_q$
irreps ($v_1$ and $v_1'$). Luckily the resulting representations
are irreducible not like at the decomposition of $t_1 \re$; when we change 
the $[k-2,2]$ factor to the trivial representation of $S_p \wr S_q$ and
take a look at the result
\beq
(\mbox{trivials} \ \ex S_p \wr S_q) \sharp \ \mbox{trivials} 
\ \otimes [1][l-1] \ex S_p \wr S_q \wr S_l
\eeq
we notice the the base representation is the identity so the 
inertia factor 
should be the full $S_l$ and not $S_1 \times S_{l-1}$ as 
considered. Thus we have to find a decomposition for the $[1] \sharp [l-1] \ex S_l$ 
representation of the inertia factor to irreps of $S_l$. It's dimension is $l$ so it may 
consist of $l$ copies of one-dimensional irreps or one vector-dimensional and 
one one-dimensional irrep. The decision is made again by evaluating characters 
on the two particular elements $(1 \ 2)$ and $(1 \ 2 \ 3)$. Finally we have: 
$s_2 = v_1' \oplus s$. 
The definitions of $v_2 v_1' \oplus v_2' v_1'$ are 
\begin{eqnarray}
v_2 v_1' &=& \langle v_1,[1] \rangle \sharp \langle s, [l-1,1] \rangle, \cr \cr
v_2' v_1' &=& \langle v_1',[1] \rangle \sharp \langle s, [l-1,1] \rangle.
\end{eqnarray}

At the $R=2$ level $v_1^2 \re$ is the only representation where we cannot use the
above mentioned method: there are two non-trivial factors in the base representation
so we cannot omit the overline and use the distributivity of the tensor product. 
So we apply the procedure similar to the one used at the $R=1$ case and obtain the
result 
\begin{eqnarray}
v_2 \stackrel{1}{\bowtie} v_2  &=& \langle v_1, [2] \rangle \sharp 
  \langle s, [l-2] \rangle, \cr \cr
v_2 \stackrel{1}{\bowtie} v_2'  &=& \langle v_1, [1] \rangle \sharp \langle v_1', [1] \rangle 
  \sharp \langle s, [l-2] \rangle, \cr \cr
v_2' \stackrel{1}{\bowtie} v_2' &=& \langle v_1', [2] \rangle \sharp \langle s, [l-2] \rangle. 
\end{eqnarray}

To conclude, let's give the precise definition of the representations 
relevant to our work. To do this, we shall define inductively certain
irreps of the multiple wreath-product
$S_{n_0}\wr S_{n_1}\wr\dots\wr S_{n_R}$. We define the representations
$s_0,v_0$ and $t_0$ of $S_{n_0}$ as
\beq s_0=[n_0]\qquad v_0=[n_0-1,1]\qquad t_0=[n_0-2,2]\eeq
We then define irreps $s_i,v_i$ and $t_i$ of $S_{n_0}\wr\dots S_{n_i}$
via the inductive rule
\begin{eqnarray}
s_{i+1}&=&\langle s_i,[n_{i+1}]\rangle\cr\cr
v_{i+1}&=&\langle v_i,[1]\rangle\sharp\langle s_i,[n_{i+1}-1]\rangle \cr \cr 
t_{i+1} &=& \langle t_i, [1] \rangle \sharp\langle s_i, [n_{i+1}-1] \rangle
\end{eqnarray}
Note that $s_i$ is just the trivial representation for all $i$, so we
can safely omit the subscript and refer to it simply as $s$.

We also need some other types of irreps, which may be constructed
starting from the representations $v_1',t_1',v_1 v_1'$ and $v_1^2$ of
$S_{n_0}\wr S_{n_1}$ defined as
\begin{eqnarray}
v_1' &=& \langle s_0,[n_1-1,1]\rangle \cr\cr
t_1' &=& \langle s_0, [n_1-2,2] \rangle \cr\cr
v_1 v_1' &=& \langle v_0, [1]\rangle\sharp\langle s_0, [n_1-1,2]\rangle\cr\cr
v_1^2 &=& \langle v_0, [2] \rangle \sharp \langle s_0, [n_1-2] \rangle 
\end{eqnarray}
The inductive step then reads
\begin{eqnarray}
v_{i+1}'&=& \langle v_i', [1]\rangle\sharp\langle s,[n_{i+1}-1]\rangle\cr\cr 
t_{i+1}'&=&\langle t_i', [1]\rangle\sharp\langle s,[n_{i+1}-1] \rangle \cr\cr 
v_{i+1}v_1'&=&\langle v_i,[1]\rangle\sharp\langle s,[n_{i+1}-2,1]\rangle\cr\cr 
v_{i+1}v_{j+1}'&=&\langle v_iv_j',[1]\rangle\sharp\langle s,[n_{i+1}-1]
\rangle\qquad\qquad j<i\cr\cr 
v_{i+1}'v_1'&=&\langle v_i',[1]\rangle\sharp\langle s,[n_{i+1}-2,1]\rangle\cr\cr
v_{i+1}'v_{j+1}'&=&\langle v_i'v_j',[1]\rangle\sharp\langle s,[n_{i+1}-1]
\rangle\qquad\qquad j<i\cr\cr
v_{i+1}^2 &=&\langle v_i^2, [1] \rangle\sharp\langle s,[n_{i+1}-1]\rangle
\end{eqnarray}
We can now define inductively all the remaining representations that we 
need  $(i > j >k)$ :
\begin{eqnarray}
v_{i+1} \stackrel{1}{\bowtie} v_{i+1} &=& \langle v_i ,
[2] \rangle \sharp \langle s, [n_{i+1}-2] \rangle \cr \cr
v_{i+1} \stackrel{k+1}{\bowtie} v_{i+1} &=& \langle v_i \stackrel{k}
{\bowtie} v_i, [1] \rangle \sharp \langle s, [n_{i+1}-1] \rangle \cr \cr
v_{i+1} \stackrel{1}{\bowtie} v_{j+1}'     &=& \langle v_i, [1] \rangle 
\sharp \langle v_j', [1] \rangle  \sharp \langle s,[n_{i+1}-2] \rangle \cr \cr 
v_{i+1} \stackrel{k+1}{\bowtie} v_{j+1}' &=& \langle v_i \stackrel{k}
{\bowtie} v_j', [1] \rangle   \sharp \langle s, [n_{i+1}-1] \rangle \cr \cr
v_{i+1}' \stackrel{1}{\bowtie} v_{j+1}'     &=& \langle v_i', [1] \rangle 
\sharp \langle v_j' ,[1] \rangle  \sharp \langle s,[n_{i+1}-2] \rangle \cr \cr 
v_{i+1}' \stackrel{k+1}{\bowtie} v_{j+1}' &=& \langle v_i' \stackrel{k}
{\bowtie} v_j', [1] \rangle   \sharp \langle s, [n_{i+1}-1] \rangle 
\end{eqnarray}

\end{document}